\documentclass[preprint,12pt]{elsarticle}

 \usepackage{epsfig}

\usepackage{amssymb}
 \usepackage{amsthm}

 \journal{Journal of Molecular Biology}

\begin{document}

\begin{frontmatter}



\title{Geometric phase shifts in biological oscillators}


\author{David S. Tourigny}

\address{MRC Laboratory of Molecular Biology, Cambridge CB2 0QH, UK}

\begin{abstract}
Many intracellular processes continue to oscillate during the cell cycle. Although it is not well-understood how they are affected by discontinuities in the cellular environment, the general assumption is that oscillations remain robust provided the period of cell divisions is much larger than the period of the oscillator. Here, I will show that under these conditions a cell will in fact have to correct for an additional quantity added to the phase of oscillation upon every repetition of the cell cycle. The resulting phase shift is an analogue of the geometric phase, a curious entity first discovered in quantum mechanics. In this Letter, I will discuss the theory of the geometric phase shift and demonstrate its relevance to biological oscillations. 
  
\end{abstract}


\end{frontmatter}


\section{Introduction}
Rhythmic cycles of gene expression underpin oscillatory processes that occur in biology with periods ranging from several years to a fraction of a second \cite{Goldbeter96}. At the cellular level, oscillatory phenomena are controlled by molecular regulators that form a network of positive or negative feedback loops (gene circuits). Positive and negative regulators, which increase and decrease gene expression respectively, are usually protein factors whose activities are in turn regulated during transcription, or at a later, post-translational stage. For example, transcription factors  CLOCK and BMAL1 regulate the levels of mRNA in the mammalian circadian clock \cite{Reppert02}, and the E3 ubiquitin ligase Mdm2 controls oscillations of the tumour suppressor p53 \cite{Lahav04}. Computational and mathematical methods have been used to study these mechanisms (reviewed in \cite{Goldbeter02}), since understanding how biological oscillations function on the molecular scale is essential for explaining the dynamics of a cell. In addition, today's research needs to address how the loss of circadian control contributes to disease at the level of an organism.  

In a recent article \cite{Gonze13}, Gonze questioned the robustness of molecular oscillations that occur concomitantly with the cell cycle. It was pointed out that most circadian clock and gene circuit models do not satisfactorily account for discontinuities in the cellular environment, since biological oscillations must also perpetuate across repetitive cell divisions \cite{Elowitz00,Mihalcescu04}. Adopting a numerical approach, Gonze demonstrated the influence of cell cycle-related effects on two popular non-linear oscillators, the Repressilator model \cite{Elowitz00} and the Goodwin model \cite{Goodwin65}. He found that although robustness diminishes for smaller periods of the cell cycle, oscillations remain relatively robust provided that the period of the cell division is much larger than the period of the oscillator \cite{Gonze13}.    

It is the purpose of this Letter to describe an effect that manifests itself on a clock (from now on `clock' will refer to any general circadian clock or gene circuit) exactly when the oscillation is considered most robust by the analysis of Gonze. More precisely, an effect that arises when the period of the cell cycle is large compared to the period of oscillations, and the cellular environment changes \emph{adiabatically} with respect to the molecular components of the clock. Under these conditions, a classical analogue of the quantum geometric phase, Hannay's angle, may be realised in a given clock system and require the cell to correct for an additional quantity added to the phase of oscillation upon every repetition of the cell cycle. Here, I will discuss the theory of geometric phase shifts and their relevance to biological systems, suggest under what conditions they may be detected, and derive Hannay's angle for two different versions of the Goodwin model.                  

\section{The classical geometric phase shift} 
Existence of the geometric phase shift in quantum mechanics was first noted by Berry \cite{Berry84} and almost immediately realised to be a \emph{holonomy} also present in other dynamical systems \cite{Simon83,Berry85,Hannay85}. A holonomy is an intrinsic property associated with any curved space, the classical example being the holonomy of the unit sphere. This holonomy is realised if one is to take a vector tangential to the sphere at a given starting point (think of a pen held on the surface of a volleyball) and then transport it around a closed loop on the surface, keeping the vector parallel to the direction of transport at every point. After completing the closed path, the vector returns to its original point, but will be rotated with respect to the direction it was pointing at the beginning of the journey. The angle of rotation is proportional to the area of the surface bounded by the path and scales with the size of the loop. It does not depend on the time taken to complete the cycle.

In Hamilton's formulation of conservative mechanics, the equations of motion describing the time evolution of a system are derived from the Hamiltonian $H$. This is a function of generalised coordinates $Q$, momenta $P$, and some constant parameters denoted by $R$. Oscillatory systems trace out an ellipse of area $2\pi I$ in $(Q,P)$ phase space, and so it is convenient to make a canonical change of coordinates to action-angle variables $(I,\theta)$ so that the equations of motion become
\begin{equation}   \frac{d \theta}{dt} = \frac{\partial H}{\partial I}= \mbox{const} = \omega_0 \ , \ \ \ \ \frac{dI}{dt} = -\frac{\partial H}{\partial \theta} = 0 \ . \end{equation}
Action-angle variables are particularly useful because frequencies $\omega_0$ of the oscillation can be obtained without ever having to solve the equations of motion. 

Hannay \cite{Hannay85} asked what would be the effect of making the parameters dependent on time, so that the vector $R$ is slowly transported around a closed loop in parameter space (slow with respect to the period of oscillations). By the assumption of adiabaticity, the period $\tau: R(t)=R(t+\tau)$ would be much larger than the period of a single orbit in $(Q,P)$ phase space, and although the path changes as the parameters are varied, the area $I$ enclosed by that path would remain the same. It turns out that after such a time $\tau$, the angle variable $\theta$ is given by the anticipated dynamical term (arising from the fact that $\theta$ is continually making orbits around the curve in phase space) plus an additional term $\Delta \theta$ depending only on the circuit in parameter space and not the duration of the process
\begin{equation} \theta(\tau) = \theta(0) + \int_0^\tau \omega_0 dt + \Delta \theta \ . \end{equation}  
For an adiabatic excursion, $dI/dt=0$, but now the equation of motion for $\theta$ is given by
\begin{equation}  \frac{d \theta}{dt} = \frac{\partial H}{\partial I} +  \frac{dR}{dt} \left\langle\frac{\partial H}{\partial R} \right\rangle \ , \end{equation}  
where the angled brackets denote the contained quantity averaged over a single period. Consequently, Hannay's angle is given by
\begin{equation} \Delta \theta = \int_0^\tau \frac{dR}{dt}  \left\langle\frac{\partial H}{\partial R} \right\rangle  dt = \oint \left\langle\frac{\partial H}{\partial R} \right\rangle dR \ . \end{equation}

The fact that the additional phase angle $\Delta \theta$ had lain undiscovered in classical mechanics for more than a century came as a great surprise to modern physicists. Together with Berry's phase it arises as a purely geometric effect of making a non-trivial loop in parameter space and is closely related to the example of the sphere described above. Shortly after its discovery, Kepler and Kagan demonstrated that time-independent geometric phase shifts also occur in dissipative systems, such as the Belousov-Zhabotinsky chemical reaction, which cannot be described by a Hamiltonian \cite{Kepler91,Kagan91}.

Realising geometric phase shifts are present in dissipative systems has deep implications for biology, which by its very nature is a complicated chemical process operating far from equilibrium. The geometric phase shift would become relevant to a biological oscillator if there exists a mechanism that transports parameters describing the cellular environment around a closed loop in parameter space. Remarkably well-suited to this task, the cell cycle provides a natural way in which the environment changes adiabatically before returning to an initial state after each cell division event. Every repetition of the cell cycle causes variations in degradation, transcription and translation rates (usually assumed to be constant in oscillator models) that could give rise to a geometric phase shift in the oscillations of a molecular clock. In the next section I will demonstrate this to indeed be the case.

\section{Geometric phase shifts induced by the cell cycle}
In the first half of this section I will derive an exact expression for Hannay's angle corresponding to a simple version of the Goodwin model. In doing so, one finds an interesting relationship to be satisfied between expression and degradation rates when $\Delta \theta$ is to contribute to the phase of an oscillation. In the second half I will consider a more complicated Goodwin model involving protein-protein interactions, for which the existence of a geometric phase shift will be demonstrated through numerical simulation. This second Goodwin model cannot be described by a Hamiltonian, and is therefore an example of a dissipative process common to many biological systems.        

Goodwin \cite{Goodwin65} proposed several models for different biological oscillators, the simplest of which can be described by a Hamiltonian $H$, a function of mRNA concentration $X$ and protein concentration $Y$. The linearised version of this model is 
\begin{equation} \frac{dX}{dt} = -\frac{\partial H}{\partial Y}= \frac{a}{A}(1-kY) - b \ , \ \ \ \  \frac{dY}{dt} = \frac{\partial H}{\partial X}=\alpha X - \beta \ , \end{equation}  
where the degradation rates $b,\beta$ and expression rates $a,A,k,\alpha$ are understood to make up a set of constant parameters $R$ on which $H$ depends. Goodwin used this system of equations to describe a closed negative feedback loop that exhibits oscillatory behaviour under the correct choice of $R$. 

To account for cell cycle effects in the linear Goodwin model it is necessary to make degradation and expression rates vary periodically in time. That is, $R(t)=R(t+\tau)$, where $\tau$ is the time taken to complete one round of the cell cycle. This means the equations become notoriously difficult to solve for arbitrary parameters. However, after making the substitutions
\begin{equation} \alpha=\frac{1}{M}  \ ,  \ \ \ \  \frac{\alpha}{ \beta}= \mu  \ , \ \ \ \   \alpha \frac{ ak}{ A^2}=\omega^2 \ , \ \ \ \ \ \frac{a}{A}-b-\frac{d}{dt}\frac{\alpha}{ \beta}=F \ , \end{equation}
the Hamiltonian $H(t)$ transforms into
\begin{equation} H(t) = \frac{1}{2M}(X^2 + \omega^2M^2Y^2) -\left(\frac{d \mu}{dt}+F\right)Y -\frac{\mu}{M}X \ , \end{equation}
which is the Hamiltonian of a classical harmonic oscillator for which action-angle variables $(I,\theta)$ are known \cite{Song00}. A second order equation of motion (independent of $X$) can be obtained for $Y$, and is satisfied by a linear combination of a particular solution $Y_p$ and two linearly-independent solutions $Y_1,Y_2$ of the homogeneous equation. Defining $\rho= \sqrt{Y_1^2+Y_2^2}$ and $\Omega= M(\dot{Y_1}{Y_2}-Y_1\dot{Y_2} )$, where the dot denotes differentiation of a solution with respect to time, it can been shown that $(I,\theta)$ are given by the relations
\begin{equation} I = \frac{1}{2\Omega} \left[\frac{\Omega^2}{\rho^2}(Y-Y_p)^2+M\frac{d\rho}{dt}(Y-Y_p)-\rho(X-M Y_p -  \mu)^2 \right] \ ,\end{equation}
\begin{equation} \cos\theta = \sqrt{\frac{\Omega}{2I}} \frac{(Y-Y_p)}{\rho} \ , \end{equation}
and
\begin{equation}  \sin \theta = \frac{1}{\sqrt{2\Omega I}} \left[ M \frac{d\rho}{dt} (Y-Y_p) - \rho (X-M Y_p -  \mu) \right] \ . \end{equation}
After a lengthy computation, one arrives at an exact expression for Hannay's angle 
\begin{equation} \Delta \theta = -\frac{1}{\Omega} \int_0^\tau M \left(\frac{d\rho}{dt} \right)^2 dt \ . \end{equation}

It is a simple matter to confirm that Hannay's angle vanishes for an oscillator with fixed expression rates. Moreover, that $\Delta \theta$ is dependent solely on solutions to the homogeneous equation of motion and therefore independent of protein and mRNA degradation rates $\beta$ and $b$, respectively. However, whilst degradation rates have no effect on the value of $\Delta \theta$ in this example, they are responsible for moving the centre of the ellipse in $(X,Y)$ phase space. The result is an interesting relation between varying degradation rates giving rise to a phase shift on the one hand, and varying expression rates contributing to the phase shift on the other. For this system at least, simultaneous variations in both expression and degradation rates are required for a non-zero phase shift upon each repetition of the cell cycle.

Introducing non-linearity into any clock should be sufficient to generate a geometric phase shift, meaning that the linear Goodwin model is by no means unique. A more complicated of Goodwin's models involves interacting protein species $Y_1$ and $Y_2$ with respective mRNA concentrations $X_1$ and $X_2$ \cite{Goodwin63}. Time evolution is governed by the equations
\begin{equation} \frac{dX_1}{dt} = \frac{a_1}{A_1+k_{11}Y_1+k_{12}Y_2}-b_1, \ \ \ \  \frac{dX_2}{dt} = \frac{a_2}{A_2+k_{21}Y_1+k_{22}Y_2}-b_2 \ , \label{eq1} \end{equation}  
and
\begin{equation} \frac{dY_1}{dt} = \alpha_1 X_1-\beta_1 \ , \ \ \ \  \frac{dY_2}{dt} = \alpha_2 X_2 -\beta_2 \ . \label{eq2} \end{equation}  
During simulations, these equations were integrated using the fourth-order Runge-Kutta method for the first 10,000 time units with biologically sensible parameter values $a_1=a_2=360$, $b_1=b_2=5$, $A_1=36$, $A_2=43$, $k_{11}=k_{22}=1$, $k_{12}=-0.5$, $k_{21}=0.1$, $\alpha_1 =  0.5$, $\alpha_2=0.6$, $\beta_1=0.2$, and $\beta_2=0.1$. This ensured that initial transients had died out long before completing the first 31-33 oscillations with a fixed period of 305 time units. After this, parameters were slowly varied around a closed loop in parameter space.  

To vary parameters, $R$ was split into two disjoint subsets so that the path in parameter space had nontrivial curvature. Not all parameters must have necessarily been varied, but $a_1$, $a_2$, $A_1$, $k_{12}$, $k_{21}$, $\alpha_1$, $\beta_1$ were chosen to make up the subset $R_c$, and $R_s$ contained $b_1$, $b_2$, $A_2$, $k_{11}$, $k_{22}$, $\alpha_2$, $\beta_2$. Elements of $R_c$ having original constant value $r_c$ were varied according to $r_c(1+0.3-0.01\cos T)$, and elements of $R_s$ according to $r_s(1+0.05-0.03\sin T)^{-1}$, where $T$ is the fraction of the path traversed in parameter space. Although appropriate for modelling purposes, this choice is somewhat arbitrary because with little data available it is difficult to know the exact form of parameter variation to use. To calculate contributions from the geometric phase in the adiabatic limit it was appropriate to use the quantity 
\begin{equation} \Delta \phi = \frac{t^+_n-t^-_n}{2\times 305} \ ,\end{equation}
where $t_n$ is the time of the $n$th peak after the loop in parameter space is traversed once either forwards ($+$) or in reverse ($-$). This formula comes from Kepler and Kagan, and is used to eliminate the dynamic phase component that has a symmetry under time reversal \cite{Kepler91}.

\begin{figure}[h]
    \begin{center}
    \epsfig{file=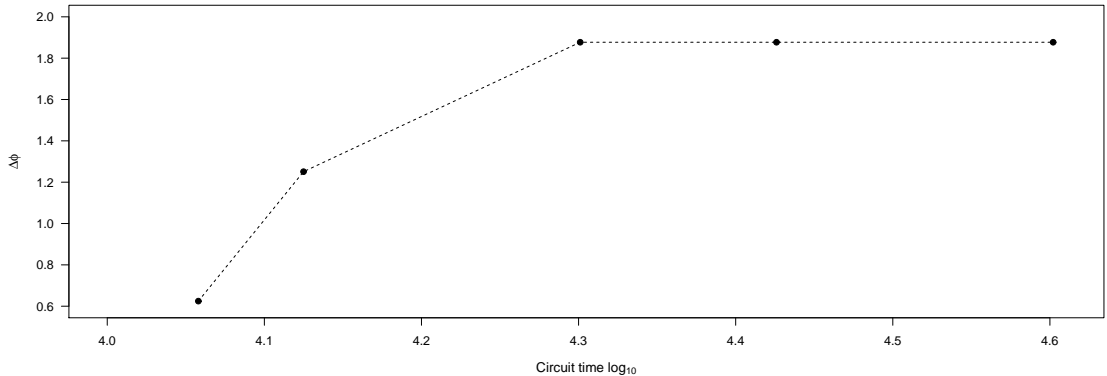,height=6cm,width=15cm}
    \end{center}
    \caption{Phase shifts for the interacting Goodwin model \cite{Goodwin63}. As described in the text, these were calculated as parameter values varied over one cycle in parameter space. When circuit time is increased they tend to the adiabatic limit of the (time-independent) geometric phase on the right.}
\end{figure}

The results of numerical simulations are displayed Fig. 1, where geometric phase shifts are seen to tend to the time-independent adiabatic limit with a value of 1.87705. The magnitude of this phase shift is comparable to that of the Belousov-Zhabotinsky reaction \cite{Kagan91}, and is for parameter values that are realistic for a plausible biological oscillator. The adiabatic limit is reached when the circuit is traversed $\sim 5 \times10^4$ times slower than the period of the oscillatory system, and so for this model the typical human cell division time of $24$ hours implies oscillations are taking place on the modest timescale of several seconds. Longer oscillations, shorter cell division times and a different choice of parameter variation will almost certainly give rise to a geometric phase shift when more complicated models with additional non-linear terms are used to describe the system. In these cases the assumption of adiabaticity is not a prerequisite for the geometric phase \cite{Aharonov87}.

\section{Concluding remarks}
In this Letter, I have shown that a classical analogue of Berry's phase has an effect on biological oscillations occurring concomitantly with the cell cycle. Perhaps the most striking result is that oscillations can experience a significant phase shift under typical growth conditions, so that a population of cells with simultaneous oscillations will quickly lose synchronisation after a few cell divisions. An organism must somehow cope with asynchrony if the regular phase is to be restored to cells, which becomes particularly important in the case of tissue-wide circadian rhythms where loss of synchronisation can lead to disease \cite{Reppert02}.    

Circadian rhythms are thought to play a role in regulating the timing and efficiency of the cell cycle, and in many tissues there is a strong correlation relating the expression of clock genes and events of cell division \cite{Matsuo03}. It therefore seems likely that cross-talk with the cell cycle is one way in which mammals prevent geometric phase shifts from destabilising the circadian clock. In addition to the circadian clock, countless other processes taking place inside growing cells display all sorts of oscillatory behaviour. These occur with periods ranging from hours to milliseconds, and so a large variety are susceptible to the geometric phase shift. 

Computational models have so far been unable to account for the high cell-to-cell variability observed in experiments with artificial gene circuits introduced into cells \cite{Elowitz00}. The geometric phase is probably a contributing factor, because artificial circuits are not coupled to any phase-correction mechanism, and are not robust to a significant change in phase. One consequence, is that the geometric phase could perhaps be realised {\em in vivo}, by introducing a synthetic gene circuit into systems developed to monitor oscillations at the single cell level \cite{Locke09}.           

I would like to acknowledge a useful discussion with G. Mitchison and J. Locke. I am funded by the Medical Research Council MC U105184332.








\end{document}